\documentclass[12pt]{article}
\newcommand{\be}{\begin{equation}}
\newcommand{\ee}{\end{equation}}
\newcommand{\bea}{\begin{eqnarray}}
\newcommand{\eea}{\end{eqnarray}}
\newcommand{\bt}{\begin{tabular}}
\newcommand{\et}{\end{tabular}}
\newcommand{\ba}{\begin{array}}
\newcommand{\ea}{\end{array}}
\newcommand{\ov}{\overline}
\newcommand{\bvec}{\mathbf}

\def\ne{\hbox{$\nu_e$ }}
\def\nm{\hbox{$\nu_\mu$ }}
\def\nt{\hbox{$\nu_\tau$ }}

\def\om{\hbox{$\omega \!\!$ }}

\def\nl{\hbox{$\nu_L \!$ }}
\def\nr{\hbox{$\nu_R \!$ }}
\def\ncl{\hbox{$\nu^c_L \!$ }}
\def\ncr{\hbox{$\nu^c_R \!$ }}

\def\rt{\hbox{$\rightarrow$ }}
\begin{document}
\setcounter{page}{0}
\thispagestyle{empty}
\baselineskip=20pt
%---------------------------------------------------------------------------

\hfill{
\begin{tabular}{l}
DSF$-$98/2 \\
INFN$-$NA$-$IV$-$98/2 \\
\end{tabular}}

\bigskip\bigskip

\begin{center}
\begin{huge}
{\bf Flavour-conserving oscillations of Dirac-Majorana neutrinos}
\end{huge}
\end{center}

\vspace{2cm}

\begin{center}
{\Large
Salvatore Esposito 
\footnote{e-mail: sesposito@na.infn.it}
\\}
\end{center}

\vspace{0.5truecm}

\normalsize
\begin{center}
{\it
\noindent
Dipartimento di Scienze Fisiche, Universit\`a di Napoli ``Federico 
II''\\
and \\
Istituto Nazionale di Fisica Nucleare, Sezione di Napoli\\
Mostra d'Oltremare Pad. 19, I-80125 Napoli Italy }
\end{center}

\vspace{3truecm}

\begin{abstract}
We analyze both chirality-changing and chirality-preserving 
transitions of Dirac-Majorana neutrinos. In vacuum, the first ones are 
suppressed with respect to the others due to helicity conservation and 
the interactions with a (``normal'') medium practically does not affect the 
expressions of the probabilities for these transitions, even if the 
amplitudes of oscillations slightly change. For usual situations 
involving relativistic neutrinos we find no resonant enhancement for 
all flavour-conserving transitions. However, for very light neutrinos 
propagating in superdense media, the pattern of oscillations 
$\nu_L \rightarrow \nu^C_L$ is dramatically altered with respect to the vacuum 
case, the transition probability practically vanishing. An application 
of this result is envisaged.
\end{abstract}

\vspace{1truecm}
\noindent

\newpage

\section{Introduction}

In this paper we correct and generalize a previous one \cite{TE} in 
which we studied Pontecorvo neutrino-antineutrino oscillations. In 
ref. \cite{TE} we confined ourselves to the zeroth order in the 
ultrarelativistic limit, which is without doubt the most significative 
one; however, from a theoretical point of view, it is interesting to 
consider also the possibility of chirality-changing oscillations 
\cite{BHP} (both in vacuum and in matter), and this will be done here 
going just a bit beyond the zeroth order. In particular in this paper 
we also correct some mistakes which brought in ref. \cite{TE} to a 
slightly incorrect formula for the survival probability in presence of 
matter oscillations (eq. (55) in \cite{TE}); however, the conclusions 
reached there remain unchanged. 

The importance of considering Dirac-Majorana neutrinos lies in the 
fact that these are described by the most general mass term in the 
electroweak lagrangian:
\be
-{\cal L}^{DM}_{m} \; = \;
\sum_{l,l^{\prime}} 
\ov{\nu}_{l^{\prime}R} 
\; M^{D}_{l^{\prime}l} \;
\nu_{l^{\prime}L} \; + \;
\frac{1}{2} \;
\sum_{l,l^{\prime}} 
\ov{\nu}^{c}_{l^{\prime}R} \; 
M^{1}_{l^{\prime}l} \;
\nu_{l L} \; + \;
\frac{1}{2} \;
\sum_{l,l^{\prime}} 
\ov{\nu}^{c}_{l^{\prime}L} 
\; M^{2}_{l^{\prime}l} \;
\nu_{l R} \; + \; h.c.
\label{11}
\ee
($l,l^{\prime} = e, \mu , \tau$), $M_D$, $M_1$, $M_2$ being the Dirac 
mass matrix and the two Majorana mass matrices, respectively. This 
scenario involves the existence of the known active states \nl, \ncr$= 
( \nl )^c$ (the superscript $c$ denotes charge conjugation) as well as 
sterile states \nr, \ncl$= ( \nr )^c$; the mass eigenstates are 
Majorana fields.\\
The mass term in (\ref{11}) is predicted in many GUTs \cite{GUT} and provides 
a simple framework for the so-called ``see-saw'' mechanism 
\cite{seesaw}, which allows to give a very small mass to neutrinos in 
a very natural way.\\
The most impressive thing related to the mass term in (\ref{11}) is, 
however, its rich phenomenology: neutrino flavour oscillations, the 
decays such as $\mu \rightarrow e \gamma$, $\mu \rightarrow 3e $, 
$\tau \rightarrow e \pi^{0}$, the conversion $\mu$-e in presence of nuclei 
$\mu^{-} + (Z,A) \rightarrow e^{-} + (Z,A)$, as well as neutrinoless 
double beta decay $(Z,A) \rightarrow (Z+2,A) + 2 e^{-}$ and 
neutrino-antineutrino oscillations can occur. All these phenomena are 
now studied experimentally \cite{PDG}.

In this paper we will concentrate our attention on the oscillation 
phenomena of Dirac-Majorana neutrinos, limiting ourselves to the case 
of flavour-conserving ones to put in evidence the salient features 
related to the non conservation of the total lepton number, which is a 
typical prediction for Majorana mass eigenstates. The most general 
case of flavour-changing oscillation phenomena will be considered 
elsewhere.\\
Since the mass term in (\ref{11}) involves both \nl, \ncr and \nr, 
\ncl, in general we can have transitions between all these states. So, 
active-active neutrino oscillations, such as \nl \rt \ncr, can occur, 
as well as active-sterile ones, such as \nl \rt \nr and \nl \rt \ncl. 
Note that even if helicity is always conserved in vacuum (this is 
related to Lorentz invariance), chirality is not in general conserved, 
and so we can have chirality-changing processes. However, this implies 
that for ultrarelativistic neutrinos ($m/k \ll 1$, where $m$ is a mass 
parameter and $k$ the neutrino momentum) propagating in vacuum the 
chirality-changing transitions are suppressed with respect to the 
other ones \cite{Prima}, because in this limit chirality is almost 
coincident with helicity, and the ``wrong'' chirality state is only a 
small component of the physical neutrino field. This suppression 
remains also for propagation in a medium, unless one considers 
helicity-flipping interactions, such as the case of neutrinos with 
magnetic moments interacting with a magnetic field \cite{mag}. In this 
work we don't deal with this last possibility, since here we will not 
consider magnetic moment interactions.\\
Our main task is to study both vacuum and matter (flavour-conserving) 
oscillations of Dirac-Majorana neutrinos, given the relevant 
importance of neutrino oscillations in astrophysics \cite{astro} and 
cosmology \cite{cosmo}. In the following section we mainly review 
vacuum oscillations (see, for example, \cite{Yana} and references 
therein) and give the expressions for the 
transition probabilities. In section 3 we develop matter oscillations 
in homogeneous media and particularly study two unusual limits, given 
the absence of resonances for flavour-conserving transitions. Finally, 
in the last section we discuss the obtained results and give our 
conclusions.

\section{Vacuum Oscillations}

Let us consider the propagation in vacuum of Dirac-Majorana neutrinos 
with 4-momentum $k^{\mu} = (\om , \bvec{k})$:
\be
{\cal L} \; = \; \ov{\nu} \, {\not k} \, \nu \; + \; \ov{\nu^c} \, 
{\not k} \, \nu^c \; - \; \frac{1}{2} \, m_D \, \left( \ov{\nu} \, \nu \; 
+ \; \ov{\nu^c} \, \nu^c \right) \, - \, \frac{1}{2} \, m_M \, \left( 
\ov{\nu} \, \nu^c \; + \; \ov{\nu^c} \, \nu \right)
\label{21}
\ee
where $m_D$, $m_M$ are the Dirac and Majorana mass, respectively. for 
the given flavour (for simplicity we assume that the mass matrices in 
(\ref{11}) are diagonal and then consider one flavour at a time). We 
take equal Majorana masses for the left-handed and right-handed 
neutrino, assuming a minimal choice for the mass parameters. In the 
chiral Weyl basis for the Dirac gamma matrices, denoting
\be
\nu \; = \; \left( \ba{c}
                   \nu_L \\
                   \nu_R    \ea \right) \;\;\;\;\;\;\;\;\;\;\;\;\;\;\;
\nu^c \; = \; \left( \ba{c}
                   \nu^c_L \\
                   \nu^c_R    \ea \right)
\label{22}
\ee
the equations of motion are given by
\be
\left( \ba{cc}
       \omega \, + \, \lambda k  &  - m_{\pm} \\
       - m_{\pm}  &  \omega \, - \, \lambda k
       \ea \right) \left( \ba{c}
                          n_{1 \pm} \\
                          n_{2 \pm}
                          \ea \right) \; = \; 0
\label{23}
\ee
where $\lambda = \pm 1$ is the helicity eigenvalue and
\be
m_{\pm} \; = \; \frac{1}{2} \, \left( m_D \, \pm \, m_M \right)
\label{24}
\ee
\bea
n_{1 \pm} & = & \pm \, \nl \, + \, \ncl
\label{25}\\
n_{2 \pm} & = & \pm \, \nr \, + \, \ncr
\label{26}
\eea
(for further reference, see \cite{TE}; note, however, that in the 
present notation \ncl = $(\nr )^c$, \ncr = $(\nl )^c$). For 
definiteness, let us consider the ``+'' states (the same will be valid 
for the ``-''states). Eq. (\ref{23}) can be written in the useful 
hamiltonian form
\be
H_+ \, \left( \ba{c}
              n_{1 +} \\
              n_{2 +}   \ea  \right) \; = \;  \om \, \left( \ba{c}
              n_{1 +} \\
              n_{2 +}  \ea \right)
\label{27}
\ee
where
\be
H_+ \; = \; \left( \ba{cc}
                   - \lambda \, k & m_+ \\
                   m_+ & \lambda \, k
                   \ea \right)
\label{28}
\ee
This hamiltonian is diagonalized by the matrix
\be
U \; = \; \left( \ba{cc}
                 \cos \, \theta_+ & \sin \, \theta_+ \\
                 - \sin \, \theta_+ & \cos \, \theta_+
                 \ea \right)
\label{29}
\ee
with
\be
\tan \, 2 \theta_+ \; = \; - \frac{m_+}{\lambda \, k}
\label{210}
\ee
and the eigenvalues are given by
\bea
E_{1+} & = & \sqrt{k^2 \, + \, m_+^2} \; \equiv \;  E_+ 
\label{211} \\
E_{2+} & = & - \, E_+
\label{212}
\eea
while the eigenvectors by
\be
\left( \ba{c}
              n_{1 +}^{\ast} \\
              n_{2 +}^{\ast}   \ea  \right) \; = \; U \; 
\left( \ba{c}
              n_{1 +} \\
              n_{2 +}   \ea  \right)
\label{213}
\ee
(the negative energy states are interpreted as usual in relativistic 
quantum theory). Similarly for the ``-'' states. So, for $\lambda \, = 
\, - 1$ the time evolution of the left-handed neutrino state is given 
by
\bea
| \nl (t)> \; = \; \frac{1}{2} \, \left( \right. & \left. \right.
\!\!\!\!\!\!\!\!\!\!\!\!\!\!  & 
\cos \, \theta_+ \, 
e^{- i E_+ t} \, |n_{1+}^{\ast}(0)> \; - \; \sin \, \theta_+ \, e^{i 
E_+ t} \, |n_{2+}^{\ast}(0)> \nonumber \\
& \!\! & \left. - \; \cos \, \theta_- \, 
e^{- i E_- t} \, |n_{1-}^{\ast}(0)> \; + \; \sin \, \theta_- \, e^{i 
E_- t} \, |n_{2-}^{\ast}(0)> \right)
\label{214}
\eea
and substituting in this expression the relation (\ref{213}) and 
(\ref{25}), (\ref{26}), we find that even if at $t = 0$ we have 
created  a \nl, at further time $t$ we can detect a \nl, as well as a 
\nr or \ncr or \ncl. The transition probabilities are given by
\bea
P(\nl \rt \nr) & = & | \, <\nr \, | \,\nl (t)> \, |^2 \; = \; 
\frac{1}{4} \, \left( \frac{m_+}{E_+} \, \sin \, E_+ t \; + \; 
\frac{m_-}{E_-} \, \sin \, E_- t \right)^2  \label{215} \\
P(\nl \rt \ncr) & = & | \, <\ncr\, | \,\nl (t)> \, |^2 \; = \; 
\frac{1}{4} \, \left( \frac{m_+}{E_+} \, \sin \, E_+ t \; - \; 
\frac{m_-}{E_-} \, \sin \, E_- t \right)^2  \label{216} 
\eea
\bea
P(\nl \rt \ncl) \; = \; | \, <\ncl \, | \,\nl (t)> \, |^2 & = & 
\sin^2 \, \frac{E_+ \, + \, E_-}{2} \, t \: 
\sin^2 \, \frac{E_+ \, - \, E_-}{2} \, t \nonumber \\
& + &
\frac{1}{4} \, \left( \frac{k}{E_+} \, \sin \, E_+ t \; - \; 
\frac{k}{E_-} \, \sin \, E_- t \right)^2  \label{217} 
\eea
while the survival probability is
\be
P(\nl \rt \nl) \; = \; 
\cos^2 \, \frac{E_+ \, + \, E_-}{2} \, t \: 
\cos^2 \, \frac{E_+ \, - \, E_-}{2} \, t \; + \;
\frac{1}{4} \, \left( \frac{k}{E_+} \, \sin \, E_+ t \; + \; 
\frac{k}{E_-} \, \sin \, E_- t \right)^2 
\label{218}
\ee
In the limit of Dirac neutrinos ($m_M = 0$), only total lepton number 
conserving processes take place \cite{Yana}
\bea
P(\nl \rt \nr) & = & \left( \frac{m_D}{2 \, E_+} \right)^2 \, \sin^2 
\, E_+ t \label{219} \\
P(\nl \rt \ncr) & = & P(\nl \rt \ncl) \; = \; 0 \label{220}
\eea
while for pure Majorana neutrinos ($m_D = 0$) we recover \cite{Yana}
\bea
P(\nl \rt \ncr) & = & \left( \frac{m_M}{2 \, E_+} \right)^2 \, \sin^2 
\, E_+ t \label{221} \\
P(\nl \rt \nr) & = & P(\nl \rt \ncl) \; = \; 0 \label{222}
\eea
Note that for the chirality-changing processes, the transition 
probabilities (\ref{215}), (\ref{216}) contain suppression factors, as 
stated in the previous section. In the ultrarelativistic limit the 
expressions for the transition probabilities simplify, and we obtain 
(for the leading terms)
\bea
P(\nl \rt \nr) & = & \frac{m_D^2}{4 \, k^2} \, \sin^2 
\, k \, t \label{223} \\
P(\nl \rt \ncr) & = & \frac{m_M^2}{4 \, k^2} \, \sin^2 
\, k \, t \label{224} \\
P(\nl \rt \ncl) & = & \sin^2 \, \frac{m_D \, m_M}{4 \, k} \, t 
\label{225}
\eea
At the leading order we thus recover the Pontecorvo formula for the 
survival probability \cite{Ponte}, \cite{TE}:
\be
P(\nl \rt \nl) \; = \; 1 \; - \; \sin^2 \, \frac{m_D \, m_M}{4 \, k} \, t 
\label{226}
\ee
Note that, apart from the suppression factors, the chirality-changing 
and the chirality-preserving transitions have very different periods, 
the oscillation length for the former being very much shorter than the 
one for the latter.\\
We also observe that for ultrarelativistic propagation the expression 
for the probability of \nl \rt \nr Dirac-Majorana neutrino 
oscillations coincides with that for Dirac neutrinos, while the 
probability for \nl \rt \ncr transitions for Dirac-Majorana neutrinos 
and pure Majorana neutrinos are also the same (cfr eqs. 
(\ref{219}),(\ref{223}) and eqs. (\ref{221}),(\ref{224})). 
This is not a real 
surprising feature, given that in the ultrarelativistic limit the 
differences between Dirac, Majorana and Dirac-Majorana neutrinos tend 
to disappear. Nevertheless, we stress that both processes can take 
place for Dirac-Majorana neutrinos and the fact that \nl \rt \nr 
transition is quite exclusively governed by the Dirac mass term, while 
\nl \rt \ncr by the Majorana mass term. Instead, both mass term must 
be non vanishing for \nl \rt \ncl oscillation to occur.

From the experimental study on disappearance experiments for neutrino 
oscillations \cite{PDG}, one can obtain the following limits for \ne 
and \nm Dirac and Majorana masses using (\ref{226}) \cite{TE}:
\be
m_D^{\nu_e} \, m_M^{\nu_e} \; \leq \; 7.5 \, \times \, 10^{-3} \; eV^2
\label{227}
\ee
and
\be
m_D^{\nu_{\mu}} \, m_M^{\nu_{\mu}} \; \leq \; 0.23 \; eV^2
\label{228}
\ee
or 
\be
m_D^{\nu_{\mu}} \, m_M^{\nu_{\mu}} \; \geq \; 1500 \; eV^2
\label{229}
\ee

\section{Matter Oscillations}

As recognized many years ago \cite{MSW}, when neutrinos propagate in a 
medium the influence of matter on neutrino oscillations can be very 
important. For example, for \ne \rt \nm oscillations (for Dirac as 
well as for Majorana neutrinos) this is due to the fact that mass 
eigenstates (which propagate as free plane waves in vacuum) do not 
coincide with flavour eigenstates (which weak-interact with the 
matter), and this difference leads to a ``resonance'' in neutrino 
oscillations, that can be reached for definite values of the squared 
masses difference between the mass eigenstates, the mixing angle and 
the matter density of the medium (for given energy of the neutrino 
beam). The presence of a resonance, obviously, completely alters the 
features of neutrino oscillations with respect to the vacuum case.\\
This is the reason of why we now study matter oscillations of 
Dirac-Majorana neutrinos; however, as we will show, for 
flavour-conserving oscillations (considered in this paper) {\it no 
resonance occurs} for ultrarelativistic neutrinos in usual conditions.

Let us start by noting that the effective potential experienced by 
neutrinos in a medium with $N_e$ electrons per unit volume and $N_n$ 
neutrons per unit volume is given by (see for example \cite{MSW}, 
\cite{Capone}, \cite{Cimento})
\bea
V_{\nl} & = & - \, b_L \label{31}\\
V_{\ncr} & = & + \, b_L \label{32}\\
V_{\nr} & = & V_{\ncl} \; = \; 0 \label{33}
\eea
where
\be
- \, b_L \; = \; \sqrt{2} \, G_F \, \left( N_e \, - \, \frac{1}{2} \, 
N_n \right)
\label{34}
\ee
for the electron flavour, while
\be
- \, b_L \; = \;  - \, \frac{G_F}{\sqrt{2}} \, N_n 
\label{35}
\ee
for the $\mu$ and $\tau$ flavours (obviously, sterile neutrinos 
experience no effective potential). Here $G_F$ is the Fermi coupling 
constant, and we consider non magnetized matter (the same is valid for 
strongly magnetized matter using for $V$ the expressions given in 
\cite{Capone}). Since the neutrino states interacting with the matter 
are the flavour eigenstates, which are linear combinations of the 
Majorana states in (\ref{25}), (\ref{26}), now the complete 
hamiltonian describing neutrino propagation in the medium cannot be 
block-diagonalized as in the vacuum case. The eigenvalue equation is 
then
\be
H \, \left( \ba{c}
              n_{1 +} \\
              n_{2 +} \\
              n_{1 -} \\
              n_{2 -}   \ea  \right) \; = \;  \om \, \left( \ba{c}
              n_{1 +} \\
              n_{2 +} \\
              n_{1 -} \\
              n_{2 -} \ea \right)
\label{36}
\ee
with
\be
H \; = \; H_{cin} \; + \; H_b
\label{37}
\ee
where
\bea
H_{cin} & = & \left(  \ba{cccc}  k & m_+ & 0 & 0 \\
                                 m_+ & - k & 0 & 0 \\
                                 0 & 0 & k & m_- \\
                                 0 & 0 & m_- & - k \ea \right)
\label{37a} \\
H_b & = & \frac{1}{2} \, \left(  \ba{cccc}  -b_L & 0 & b_L & 0 \\
                                            0 & b_L & 0 & b_L \\
                                            b_L & 0 & -b_L & 0 \\
                                            0 & b_L & 0 & b_L \ea \right)
\label{37c}
\eea
Now, to find the transition probabilities one firstly has to 
diagonalize the complete hamiltonian in (\ref{37}). This can be done 
analytically, because the eigenvalue equation corresponds to a fourth 
degree algebraic equation whose solutions in terms of radicals are 
known. However, if we do so, we will obtain ugly formulae for the 
probabilities, and the physical content of them cannot be extracted in 
an easy way. So we will now proceed to consider some physical 
approximations in the framework of which we can easily get the 
transition probabilities and discuss their implications. \\
First of all we want to analyze the problem of occurrence of resonances 
in the transition processes (for relativistic neutrinos), looking for 
possible level crossings in the flavour eigenstate basis \cite{MSW}. 
The hamiltonian, and then the eigenvalue equation, in this basis can 
be obtained from (\ref{36}), (\ref{37}) remembering the relations in 
(\ref{25}), (\ref{26}):
\be
\left(  \ba{cccc}
k & 0 & m_D / 2 & m_M / 2 \\
0 & k \, - \, b_L & m_M / 2 & m_D / 2 \\
m_D / 2 & m_M / 2 & - \, k \, + b_L & 0 \\
m_M / 2 & m_D / 2 & 0 & - \, k \ea \right) \, 
\left( \ba{c}  \ncl \\
               \nl \\
               \ncr \\
               \nr \ea \right) \; = \; \om \,
\left( \ba{c}  \ncl \\
               \nl \\
               \ncr \\
               \nr \ea \right) 
\label{38}
\ee
From this we immediately recognize that the level crossing for \nl \rt 
\ncl transition is given by the condition
\be
b_L \; = \; 0
\label{39}
\ee
while those for \nl \rt \ncr and \nl \rt \nr transitions are given, 
respectively, by
\bea
b_L & = & k  \label{310} \\
b_L & = & 2 \, k \label{311}
\eea
The last two conditions are never verified by relativistic neutrinos
\footnote{However, there is an exception offered by the cosmological 
relic neutrinos in the Universe, whose momentum  is $10^{-2} \div 10^{-4}$ eV 
\cite{Kolb}} because the effective potential for media encountered in 
Nature never exceeds some eV (this ``huge'' value holding for the very 
dense matter of a neutron star), due to the Fermi coupling constant in 
(\ref{34}), (\ref{35}). \\
The condition for \nl \rt \ncl transition 
in (\ref{39}) is not properly a ``resonance condition'', 
because it does not involve neutrino energy and mass parameters; it 
tells us, however, that the amplitude of oscillations is maximum in 
vacuum, so that the interaction of \nl with the medium can only 
suppress neutrino oscillation (as already discussed in \cite{TE}, this 
is obvious looking at eq. (\ref{225}), where we see that in vacuum the 
amplitude of oscillation is already 1). Nevertheless, the implications 
of (\ref{39}) can be important in the study of propagation of 
$\nu_{eL}$ during the neutronization phase of a supernova, where the 
condition $N_e \, = \, N_n / \! 2$ is fulfilled. \\
Then we can conclude by saying that in ``normal'' media, at least for 
relativistic neutrinos, no resonance occurs for (flavour-conserving) 
oscillations of Dirac-Majorana neutrinos.\\
With this result in mind, we now proceed to evaluate the transition 
probabilities in two unusual limits, namely for ultrarelativistic 
neutrinos satisfying the conditions
\be
{\mathrm case \; A}: \;\;\;\;\;\;\;\;\;\;\;\;\;\;\; 
b_L \; \ll \; m_{D,M} \; , k
\label{312}
\ee
or
\be
{\mathrm case \; B}: \;\;\;\;\;\;\;\;\;\;\;\;\;\;\; 
m_{D,M} \; \ll \; b_L \; , k
\label{313}
\ee
Since neutrinos will meet no resonance during their propagation in 
matter (we do not discuss here the situation envisaged above, for 
which $N_e \, = \, N_n / \! 2$) these two cases can be analyzed by 
using perturbation theory.

\subsection{Case A}

In the eigenvalue problem (\ref{36}), (\ref{37}) we consider $H_0 \, = 
\, H_{cin}$ as the unperturbed hamiltonian and $H_b$ as a 
perturbation. The unperturbed energy levels and states are given by 
(\ref{211}), (\ref{212}) and (\ref{213}), respectively. By means of 
standard techniques, we find that, at first order in $b_L$, the 
hamiltonian eigenvalues are approximatively given by
\bea
E_1 & = & - \, ( E_+ \; - \; \frac{b_L}{2} \, \cos , 2 \theta_+ ) \; 
\equiv \; - \, E_{+ m} \label{314}\\
E_2 & = & E_{+ m} \label{315}\\
E_3 & = & - \, ( E_- \; - \; \frac{b_L}{2} \, \cos , 2 \theta_- ) \; 
\equiv \; - \, E_{- m} \label{316}\\
E_2 & = & E_{- m} \label{317}
\eea
while the (approximated) matter eigenstates and flavour eigenstates 
are related in the following way:
\be
\left( \ba{c} \nu_1 \\ \nu_2 \\ \nu_3 \\ \nu_4 \ea \right) \; = \; 
\frac{1}{\sqrt{2}} \, \left( \ba{cccc}
- C_+ \, + \, D_- & - C_+ \, - \, D_- & A_+ \, - \, B_- & A_+ \, + \, 
B_- \\
A_+ \, + \, B_- & A_+ \, - \, B_- & C_+ \, + \, D_- & C_+ \, - \, D_- 
\\
- C_- \, + \, D_+ & C_- \, + \, D_+ & A_- \, - \, B_+ & - A_- \, - \, 
B_+ \\
A_- \, + \, B_+ & - A_- \, + \, B_+ & C_- \, + \, D_+ & - C_- \, + \, 
D_+ \ea \right) \, \left( \ba{c} \ncl \\ \nl \\ \ncr \\ \nr \ea 
\right)
\label{318}
\ee
with
\bea
A_+  & = & \cos \, \theta_+ \; - \; \frac{b_L}{4} \, \frac{\sin \, 
\theta_+ \, \sin \, 2 \theta_+}{E_+} \label{319} \\
B_+ & = & \frac{b_L}{2} \, \left( \frac{\sin \, \theta_+ \, \sin \, 
(\theta_+ \, - \, \theta_-)}{E_+ \, + \, E_- } \; - \; 
\frac{\cos \, \theta_+ \, \cos \, 
(\theta_+ \, - \, \theta_-)}{E_+ \, - \, E_- } \right) \label{320} \\
C_+  & = & \sin \, \theta_+ \; + \; \frac{b_L}{4} \, \frac{\cos \, 
\theta_+ \, \sin \, 2 \theta_+}{E_+} \label{321} \\
D_+ & = & - \, \frac{b_L}{2} \, \left( \frac{\cos \, \theta_+ \, \sin \, 
(\theta_+ \, - \, \theta_-)}{E_+ \, + \, E_- } \; + \; 
\frac{\sin \, \theta_+ \, \cos \, 
(\theta_+ \, - \, \theta_-)}{E_+ \, - \, E_- } \right) \label{322} \\
\eea
and similarly for $A_-$, $B_-$, $C_-$, $D_-$ by substituting ``+'' \rt 
``-''. With some calculations, one can check that the matter mixing 
matrix in  (\ref{318}) is unitary (orthogonal) at first order in 
$b_L$. Given this matrix, we are now able to calculate the transition 
probabilities for matter oscillations in the present limit. For 
simplicity, we report here only the formulae which are appropriate for 
ultrarelativistic neutrinos:
\bea
P_m(\nl \rt \nr) & \simeq & \frac{m_D^2}{4 \, k^2} \, \left( 1 \, + \, 
\frac{b_L}{k} \right) \, \sin^2 
\, k \, t \label{323} \\
P_m(\nl \rt \ncr) & \simeq & \frac{m_M^2}{4 \, k^2} \, \left( 1 \, + \, 2 
\, \frac{b_L}{k} \right) \, \sin^2 
\, k \, t \label{324} \\
P_m(\nl \rt \ncl) & \simeq & \sin^2 \, \frac{m_D \, m_M}{4 \, k} \, \left( 
1 \, + \, \frac{b_L}{2 k} \right) \, t 
\label{325} \\
P_m(\nl \rt \nl) & \simeq & 1 \, - \, P_m(\nl \rt \ncl) \, - \, 
P_m(\nl \rt \nr) \, - \, P_m(\nl \rt \ncr)
\label{326}
\eea
The last relation corrects eq. (55) of ref. \cite{TE}.\\
Confronting the obtained results with the expressions 
(\ref{223})-(\ref{226}) holding for the vacuum case, we can observe 
that for the chirality-changing transitions the period of matter and 
vacuum oscillations are the same, while their amplitude change 
(because of the sign of $b_L$, the amplitude of matter oscillations 
are always lower than those of vacuum oscillations for \nm and \nt, 
while for \ne all possibilities can occur). The opposite happens for 
\nl \rt \ncl oscillations. In any case, the corrections due to matter 
interaction are quite insignificant, because of the very smallness of 
the effective potential  with respect to neutrino momentum in normal 
situations, except for relic neutrinos propagating in very dense stars 
as in the study developed in \cite{Kiers}. However, for the last case, 
relations (\ref{323})-(\ref{326}) would not apply because of the non 
relativistic propagation; the right expressions obtained relaxing the 
assumption of ultrarelativistic propagation can be found in the 
appendix.

\subsection{Case B}

Let us now consider the situation in which the mass parameters are the 
smallest ones, namely (\ref{313}) holds. In this case, it ie more 
convenient to work in the flavour basis and then solve the eigenvalue 
equation (\ref{38}) by considering $H_0 \, = \, diag \{ k, k - b_L, 
- k + b_L , - k \}$ as the unperturbed hamiltonian and the remaining 
hamiltonian mass term in (\ref{38}) as a perturbation. At second order 
in the perturbation parameters we thus obtain the following energy 
levels:
\bea
E_1 & \simeq & k \; + \; \frac{m_D^2}{4 (2 k \, - \, b_L )} \; + \; 
\frac{m_M^2}{8 k} \label{327} \\
E_2 & \simeq & k \; - \; b_L \; + \; \frac{m_D^2}{4 (2 k \, - \, b_L )} 
\; + \; \frac{m_M^2}{8 ( k \, - \, b_L )} \label{328} \\
E_3 & \simeq & - \, E_2 \label{329} \\
E_4 & \simeq & - \, E_1 \label{330} 
\eea
Instead, at the same order, the energy eigenstates are given by
\be
\left( \ba{c} \nu_1 \\ \nu_2 \\ \nu_3 \\ \nu_4 \ea \right) \, = \, 
\left( \ba{cccc}
1 - \frac{1}{2} ( A^2 + C^2 ) & A \, C \, \delta_1 & A & C 
\\  
- \, A \, B \, \delta_2 & 1 - \frac{1}{2} ( A^2 + B^2 ) & 
B & A \\
- \, A & - \, B & 1 - \frac{1}{2} ( A^2 + B^2 ) & - \, A 
\, B \, \delta_2 \\
- \, C & - \, A & A \, C \, \delta_1 & 1 - \frac{1}{2} ( A^2 + 
C^2 )  \ea \right) \left( \ba{c} \ncl \\ \nl \\ \ncr \\ \nr \ea 
\right)
\label{331}
\ee
where
\bea
A & = & \frac{m_D}{2 ( 2 k \, - \, b_L )} \label{332} \\
B & = & \frac{m_M}{4 ( k \, - \, b_L )} \label{333} \\
C & = & \frac{m_M}{4 k} \label{334} \\
\delta_1 \; = \; \frac{4 k \, - \, b_L}{b_L} \;\;\;\;\;
& , &  \;\;\;\;\; \delta_2 \; = \; \frac{4 k \, - \, 3 b_L}{b_L} 
\label{335}
\eea
With the mixing matrix defined in (\ref{331}) we can now obtain the 
expressions for the transition probabilities
\bea
P_m (\nl \rt \nr) & \simeq & 4 \, A^2 \, \sin^2 \,  \frac{E_1 \, + \, 
E_2}{2} \, t  \label{336} \\
P_m (\nl \rt \ncr) & \simeq & 4 \, B^2 \, \sin^2 \,  E_2 \, t  
\label{337} \\
P_m (\nl \rt \ncl) & \simeq & 0 \label{338}
\eea
More interestingly, for $b_L \ll k$ we have
\bea
P_m(\nl \rt \nr) & \simeq & \frac{m_D^2}{4 \, k^2} \, \left( 1 \, + \, 
\frac{b_L}{k} \right) \, \sin^2 
\, k \, t \label{339} \\
P_m(\nl \rt \ncr) & \simeq & \frac{m_M^2}{4 \, k^2} \, \left( 1 \, + \, 2 
\, \frac{b_L}{k} \right) \, \sin^2 
\, k \, t \label{340} \\
P_m(\nl \rt \ncl) & \simeq & 0 \label{341} 
\eea
and we immediately recognize that the transition probabilities for \nl 
\rt \nr and \nl \rt \ncr are coincident with those calculated for the 
case A, this showing that the presence of a medium has practically in 
any regime no influence on these processes. Instead, the very 
intriguing fact is that the probability for \nl \rt \ncl  
substantially vanishes in the limit (\ref{313}): one can check that 
$P_m (\nl \rt \ncl)$ has in fact an oscillatory behaviour with an 
amplitude of the order $\frac{m_D^2}{k^2} \, \frac{m_M^2}{k^2}$. This 
feature is a general one of all the chirality-preserving transitions, 
and is a consequence of the fact that, in the present limit, the 
physical propagating neutrino fields are not predominantly Majorana 
states. In fact let us consider, for example, the state $\nu_2$ in 
(\ref{331}). Switching off the perturbation, this correspond to the 
flavour state \nl. Because of the mass interaction, this state 
acquires, at first order, a small \nr and/or \ncr component letting non 
zero $m_D$ and/or $m_M$ but no fraction of \ncl enters in $\nu_2$ at 
this order. Only at the second order, for Dirac and Majorana masses 
both non vanishing, $\nu_2$ acquires a \nl component; so the 
predominant flavour content of $\nu_2$ is only due to \nl but not to 
\nl and \ncl in almost equal parts as happens in vacuum (or in the 
case A), and this explains the result (\ref{341}). Analogous 
considerations hold for the other chirality-preserving transitions.

\section{Discussion and conclusions}

In this paper we have studied flavour-conserving oscillations of 
Dirac-Majorana neutrinos when these propagate in vacuum as well as in 
a dense medium. \\
In vacuum, since the physical neutrino field of definite helicity is 
predominantly composed of states with a given related chirality and 
has only a small component (for non zero masses) of states with the 
other chirality, both chirality-changing and chirality-preserving 
transitions are possible, but the first ones are suppressed with 
respect to the others \cite{Prima}. In the ultrarelativistic limit, 
\nl \rt \nr oscillations are ruled exclusively by the Dirac mass term, 
and \nl \rt \ncr ones exclusively by the Majorana mass term, while \nl 
\rt \ncl transitions can take place only if both $m_D$ and $m_M$ are 
non zero (more precisely, the Dirac mass term is essential for the 
existence of the sterile \ncl state, while the total lepton number violating 
Majorana mass term has to be present for the $\Delta \, L \, = \, 2$ 
process \nl \rt \ncl to occur).\\
When energetic neutrinos propagate in a medium, the oscillations 
considered here undergo no resonant enhancement since the effective 
potential experienced by neutrinos is too low with respect to neutrino 
momentum, and this justifies a perturbative analysis of the phenomena, 
as made in the present paper. The chirality-changing transitions 
proceed in matter practically as in vacuum, apart from an 
insignificant modification of the oscillation amplitude. For the \nl 
\rt \ncl transitions, if the mass interaction is more effective than 
the matter interaction (case A, (\ref{312})), the period of 
oscillation slightly changes with respect to the vacuum case, but even 
here the correction is unimportant because of the smallness of the 
effective potential. Instead the situation dramatically change if the 
interaction of neutrinos with matter is predominant on mass 
interaction. Due to the fact that, in this case, the physical 
propagating neutrino field is no longer dominated by the vacuum 
Majorana mass states (as in the previous case), the probability for 
\nl \rt \ncl transitions is practically zero. This result would then 
suggest a peculiar situation in which one can study neutrino 
oscillation phenomena by using the fact that $P(\nl \rt \ncl)$ in 
(\ref{225}) has an oscillatory behaviour in time (or distance) while 
$P_m (\nl \rt \ncl)$ in (\ref{341}) is {\it de facto} time 
independent. Indeed, let us suppose that at $x \, = \, 0$ a pure \nl 
beam is produced and at $x \, = \, L$ a neutrino detector is placed. 
If between the source and the detector there is the vacuum, then at $x 
\, = \, L$ we have a probability to detect \ncl given by (\ref{225}) 
with $t \, \simeq \, x \, = \, L$. The situation change if between the 
source and the detector we place a dense medium from $x_1$ to $x_2$ 
(with $x_2 \, - \, x_1 \, = \, L_m$) with a density satisfying 
(\ref{313}): in this case, because of (\ref{341}), the probability of 
finding a \ncl at $x_1$ and at $x_2$ is the same, and the effective 
length for oscillation is no longer $L$ but $L \, - \, L_m$. So, we 
can detect at $x \, = \, L$ a \ncl with a probability again given by 
(\ref{225}) but with $t \, \simeq \, x \, = \, L \, - \, L_m$. This 
method would be useful to study oscillations of neutrinos with a very 
small mass (the relation (\ref{313}) has to be satisfied); however, we 
think that it is unfair to apply this method in a very near future. 
Nevertheless, the results contained in this paper could be interesting 
for astrophysics and especially cosmology, where very dense matter 
indeed exists.

\vspace{1truecm}

\noindent
{\Large \bf Acknowledgements}\\
\noindent
I gratefully acknowledge Dr. N. Tancredi for useful comments and his
encouragement.

\newpage

\noindent
{\Large \bf Appendix}\\

\begin{appendix}

From (\ref{318}), assuming no hierarchy of $m_D$, $m_M$ with respect 
to neutrino momentum $k$, we obtain the following expressions for the 
transition probabilities:
\bea
P_m (\nl \rt \nr) & \simeq &
\frac{1}{4} \, \left( \frac{m_+}{E_+} \, \sin \, E_{+m} t \; + \; 
\frac{m_-}{E_-} \, \sin \, E_{-m} t \right)^2 
\; + \; \frac{b_L}{4} \, \left( \frac{m_+}{E_+} \, \sin \, E_{+m} t \; + 
\right. \nonumber \\ & + & \left.
\frac{m_-}{E_-} \, \sin \, E_{-m} t \right) \, \left(
\frac{m_+ \, k}{E_+^3} \, \sin \, E_{+m} t \; + \; 
\frac{m_- \, k}{E_-^3} \, \sin \, E_{-m} t \right)
\eea
\bea
P_m (\nl \rt \ncr) & \simeq &
\frac{1}{4} \, \left( \frac{m_+}{E_+} \, \sin \, E_{+m} t \; - \; 
\frac{m_-}{E_-} \, \sin \, E_{-m} t \right)^2 
\; + \; \frac{b_L}{2} \, \left( \frac{m_+}{E_+} \, \sin \, E_{+m} t \; +
\right. \nonumber \\ & - & \left. 
\frac{m_-}{E_-} \, \sin \, E_{-m} t \right) \, \left(
\frac{m_+ \, k}{2 \, E_+^3} \, \sin \, E_{+m} t \; - \; 
\frac{m_- \, k}{2 \, E_-^3} \, \sin \, E_{-m} t \; + 
\right. \nonumber \\ & + & \left. 
\frac{m_+ \, - 
\, m_-}{E_+^2 \, - \, E_-^2} \, \frac{k}{E_+ \, E_-} \, \left( E_+ \, 
\sin \, E_{- m} t \; - \; E_- \, \sin \, E_{+ m} t \right) \, 
\right)
\eea
\bea
P(\nl \rt \ncl) & \simeq &
\sin^2 \, \frac{E_{+m} \, + \, E_{-m}}{2} \, t \: 
\sin^2 \, \frac{E_{+m} \, - \, E_{-m}}{2} \, t \; + \;
\frac{1}{4} \, \left( \frac{k}{E_+} \, \sin \, E_{+m} t \; +
\right. \nonumber \\ & - & \left.
\frac{k}{E_-} \, \sin \, E_{-m} t \right)^2 
\; + \; \frac{b_L}{2} \, \left( \frac{k}{E_+} \, \sin \, E_{+m} t \; - \; 
\frac{k}{E_-} \, \sin \, E_{-m} t \right) \, \times \nonumber \\
& \times &
\left(
\frac{m_-^2}{E_-^2} \, \frac{\sin \, E_{-m} t}{2 \, E_-} \; - \; 
\frac{m_+^2}{E_+^2} \, \frac{\sin \, E_{+m} t}{2 \, E_+} \right)
\eea
In the limit $m_D$, $m_M$ $\ll$ $k$ these expressions reduce to those 
reported in eqs. (\ref{323})-(\ref{325}).

\end{appendix}

\end{document}